\documentclass[journal]{IEEEtran}

\IEEEoverridecommandlockouts                              

\ifCLASSINFOpdf
  \usepackage[pdftex]{graphicx}
  \graphicspath{{./Figs/}{./Figs/}}
  \DeclareGraphicsExtensions{.pdf,.jpeg,.png}
\else
   \usepackage[dvips]{graphicx}
   \graphicspath{{./Figs/}}
   \DeclareGraphicsExtensions{.eps}
\fi

\usepackage{booktabs,array}

\newcommand{\tableequation}[1]{%
	\vspace*{-\baselineskip}
	{\begin{flalign}#1&&&\end{flalign}}%
	\vspace*{-\baselineskip}
}

\usepackage{amsmath} 
\usepackage{amssymb}  
\usepackage{array}
\usepackage{amsthm}
\usepackage{rotating}
\usepackage{rotfloat}
\usepackage{enumitem}
\usepackage[export]{adjustbox}
\usepackage{arydshln}
\usepackage{float}
\usepackage{tabularx}
\usepackage{tabularx}
\usepackage{multirow}
\usepackage[numbers,sort&compress]{natbib}
\usepackage{amsmath}
\usepackage{enumitem}
\usepackage{bm}
\usepackage{textcomp}
\usepackage{url}

\newcommand{\diag}{\mathop{\mathrm{diag}}}

\usepackage{booktabs}
\newcommand{\tabitem}{~~\llap{\textbullet}~~}
\usepackage{tabularx}
\usepackage{amsmath}

\title{Interpolation of Sparse Graph Signals by Sequential Adaptive Thresholds}
\author{$^\ast$ 
	Mahdi Boloursaz Mashhadi, \textit{Student Member, IEEE,} Maryam Fallah and Farokh Marvasti, \textit{Senior Member, IEEE}
	
	\thanks{$^\ast$ The authors are with the Advanced Communications Research Institute (ACRI), EE Department, Sharif University of Technology (SUT), Tehran, Iran. (email: boloursaz@ee.sharif.edu,  maryamfallah1994@gmail.com, marvasti@sharif.edu )}
	\thanks{}}

\begin{document}

\maketitle \thispagestyle{empty} \pagestyle{empty}

\begin{abstract}
This paper considers the problem of interpolating signals defined on graphs. A major presumption considered by many previous approaches to this problem has been low-pass/band-limitedness of the underlying graph signal. However, inspired by the findings on sparse signal reconstruction, we consider the graph signal to be rather sparse/compressible in the Graph Fourier Transform (GFT) domain and propose the Iterative Method with Adaptive Thresholding for Graph Interpolation (IMATGI) algorithm for sparsity promoting interpolation of the underlying graph signal. We analytically prove convergence of the proposed algorithm. We also demonstrate efficient performance of the proposed IMATGI algorithm in reconstructing randomly generated sparse graph signals. Finally, we consider the widely desirable application of recommendation systems and show by simulations that IMATGI outperforms state-of-the-art algorithms on the benchmark datasets in this application.
\end{abstract}
\section{Introduction}
\label{sec:intro}
Interpolating signals defined on graphs is a basic problem  that has found numerous applications in a variety of fields such as sensor networks, data classification, brain-imaging and recommendation systems \cite{hagmann2008mapping, sandryhaila2013discrete, sandryhaila2014discrete,hoche2008fast} The aim of interpolation is to find missing values of a graph signal from its values on a subset of the nodes assuming a particular signal model e.g. band limitedness in the GFT (Graph Fourier Transform) domain, Sparsity, and etc. Different algorithms have been proposed for this problem so far \cite{narang2013signal, narang2013localized,segarra2015interpolation,belkin2004semi,chen2009fast,grady2010anisotropic} (See \cite{shuman2013emerging}for an extensive review).

The K-Nearest Neighbor (KNN) method proposed in \cite{chen2009fast} is a basic technique that uses an efficient Lanczos procedure for recursive data partitioning and reconstructs the unknown signal values using  a weighted combination of  the known values on the k-nearest nodes \cite{grady2010anisotropic}. It is known that KNN overlooks the dependencies existing between the known samples. However, more computationally demanding algorithms have been proposed by \cite{narang2013signal,belkin2004semi,grady2010anisotropic} that take more similarity factors into account and thus provide more accurate estimates. 

The method proposed by \cite{narang2013localized} shows improved performance regarding both accuracy and computational efficiency and serves as a benchmark for performance comparisons in this research. \cite{narang2013localized} proposes a Regularization Based Method (RBM) in order to minimize a cost function consisting of both signal smoothness and the square reconstruction error. Furthermore, it proposes the Iterative Least Square Reconstruction (ILSR) method for graph signal reconstruction based on band-limitedness. \cite{narang2013localized} also provides a comprehensive comparison between the performance of state-of-the-art interpolation methods for the application of recommendation systems working on three benchmark datasets of Movielens \cite{WinNT}, Jester \cite{WinNT2} and Books \cite{WinNT3}.

The Iterative Weighting Reconstruction (IWR) and Iterative Propagation Reconstruction (IPR) methods were proposed by \cite{wang2015local} to reconstruct band-limited graph signals by the idea of division to sub-graphs. Compared to ILSR in \cite{narang2013localized}, these methods achieve improved convergence rates, however the partitioning technique creates isolated local-sets which leads to reluctant sampling vertices. 

In \cite{shuman2013emerging}, a three layer cluster division is proposed which is similar to \cite{wang2015local} but reduces the sampling rate by removing the isolated vertex sets.

\textbf{\textit{Contributions:}} As observed above, a major presumption that has been considered in many previous works on graph signal interpolation \cite{narang2013signal,narang2013localized,segarra2015interpolation,shuman2013emerging,anis2015asymptotic} is that the signal defined on the graph is band-limited and there are a few prior works that assume sparsity \cite{zhu2012graph,marques2016sampling}. In this work, we consider the graph signal to be sparse/compressible rather than band-limited in the GFT domain i.e. it has a few non-negligible coefficients spread along the whole GFT range without prior knowledge of their locations. We propose the Iterative Method with Adaptive Thresholding for Graph Interpolation (IMATGI) for sparsity promoting reconstruction of graph signals. We provide the convergence analysis for the proposed method and show its efficient performance by simulations. Another key contribution of this work is that we show (by extensive simulations on the benchmark datasets used by \cite{narang2013signal,narang2013localized},\cite{hofmann2005collaborative,ziegler2005improving,goldberg2001eigentaste} that applying the sparse signal assumption by IMATGI significantly improves the interpolation performance in the widely desirable application of recommendation systems. This observation brings us to the conclusion that the natural Movies, Jokes, Books and etc. datasets better match the sparse signal assumption rather than the classic band-limitedness.

\textbf{\textit{Notations}}: Throughout this paper, we denote scalar values and vectors by italic and regular lowercase letters, respectively. Matrices and sets are denoted by boldface and regular uppercase letters. Finally, calligraphic letters denote mathematical operators and $E\{.\}, (.)^t$ and $||.||_2$ are expected value, matrix transposition and the second vector norm, respectively.

The rest of this paper is organized as follows. Section \ref{sec:matmodel} introduces the proposed IMATGI algorithm. Section \ref{sec:converge} analytically discusses the reconstruction capability of IMATGI. Section \ref{sec:sim} includes the simulation results and performance comparisons and finally section \ref{sec:con} concludes the paper.

For further reproduction of the reported simulation results, MATLAB codes are made available on $ee.sharif.edu/\sim boloursaz$.

\section{The proposed IMATGI algorithm}

\label{sec:matmodel}
In this subsection, we present the proposed Iterative Method with Adaptive Thresholding for Graph Interpolation (IMATGI) algorithm. This algorithm assumes that the underlying graph signal is sparse/compressible in the Graph Fourier Domain (GFT) and gradually extract the significant signal components by iterative thresholding of the estimated signal with a decreasing threshold. This technique is inspired by the previous findings on sparsity promoting reconstruction of regular signals from missing samples \cite{marvasti2012unified}. 

Consider an undirected graph $G=(V,E)$ with $V$ as the set of vertices and $E$ as edges. Denote by $\textbf{L}$ the symmetric normalized Laplacian matrix for this graph as defined by \cite{narang2013localized}. Now, decompose $\mathbf{L=U\Lambda{U}^t}$ in which $\mathbf{\Lambda}=\diag({\mathit{\lambda}_{1},\mathit{\lambda}_{2},\cdots,\mathit{\lambda}_{n}})$ is a diagonal matrix of non-negative eigenvalues and $\mathbf{U}=[{u}_1,{u}_2,\cdots,{u}_n]$ is a unitary matrix containing the corresponding eigenvectors. 

Now, define the corresponding graph signal as a function $\text f:V\rightarrow R$ and denote it by the vector $\text f\in R^N $where the $i$th component represents the signal value on the $i$th vertex. Considering the eigenvectors $u_i$ as the basis vectors and the corresponding eigenvalues $\lambda_i$ as the graph frequencies (as defined by [12]), this signal can be transformed into the Graph Fourier Transform (GFT) domain by ${\hat{{\text f}}=\textbf{U}^t\text f}$. 

In the graph interpolation problem, the signal entries are known only on a subset of nodes $\textbf{S}$ and we aim to interpolate the unknown signal values on $\textbf{S}^c$. Define the diagonal sampling matrix $\mathbf{S_{N\times N}}=\diag(s_{1},s_{2},\cdots,s_{N})$ in which $i$th diagonal element is defined by: 
\begin{eqnarray}
\label{eq:sol_1}
s_i= \bigg\{{\genfrac{}{}{0pt}{}{1\quad if\quad i\in \textbf{S}}{0\quad if \quad i\not\in \textbf{S}}} 
\end{eqnarray}

Hence, the sub-sampled signal is given by $\text f_s=\textbf{S}\text f$. Utilizing this notation, the proposed IMATGI algorithm is presented in Table~\ref{tab:t1}. The idea of reconstructing sparse graph signals by sequential thresholdings according to the IMATGI update rule (\ref{eq:sol}) is a rational guess that is inspired by the Iterative Hard Thresholding (IHT) \cite{IHT1} and Iterative Method with Adaptive Thresholding (IMAT) \cite{marvasti2012unified} algorithms in the literature of sparsity promoting reconstruction of regular signals from missing samples.

In  Table~\ref{tab:t1}, $\text f$ and $\text f_k$ denote the original signal and its reconstructed version at the $k$th algorithm iteration. $\mathit{\lambda}$ is the relaxation parameter that controls the convergence rate of the algorithm and $\mathcal T(.)$ denotes the thresholding operator. 
The thresholding block operates elementwise on the input vector and sets the vector entries with absolute values below the threshold to zero. The threshold value $t(k)$ is decreased exponentially by $t(k)=\beta e^{-\alpha k}$ where $k$ is the iteration number. The algorithm parameters $\lambda$,$\beta$,$\alpha$ are optimized empirically for fastest convergence. 
\begin{table}[H]
		\centering
		\caption {Stepwise presentation for IMATGI  } \label{tab:t1}
		\begin{tabular}{|l|l|}
			\hline
			\multicolumn{1}{|c|}{IMATGI Algorithm for Sparse Signal Reconstruction on Graphs} \\
			\hline \\
	    	\textbf	{Require: $G0 = (V,E)$} \\
	    	\quad\tabitem{Compute normalized Laplacian matrix} \\	
			\textbf{Inputs}: \\
			 \quad\tabitem 	$\mathbf{S_{N\times N}}$: The sampling matrix\\
			 \quad\tabitem \textit{$\epsilon$}: Stopping criteria\\
			 \quad\tabitem $\mathbf{(\alpha,\beta,\lambda)}$: Algorithm Parameters \\ 
		
			\textbf{Output}: \\ 
			 \quad\tabitem  {$ {\tilde{\text{f}}_{N\times 1}}$}: The reconstructed signal\\
			\textbf{Algorithm}: \\
			 \quad\tabitem Initialization  {$\tilde{\text{f}} =\text f_1=\textbf{S}\text {f},\text {f}_0=0_{N \times 1}, k=1$}\\
			 \quad\tabitem While $(||\text{f}_k-\text{f}_{k-1}||>\epsilon)$  \\
			\quad \quad  \quad  -~Calculate the threshold as: $t(k)=\beta e^{-\alpha k}$ \\
			\quad \quad  \quad  -~Perform the thresholding as:  $\text{g}_k=\textbf{U}(\mathcal{T}(\textbf{U}^t\text{f}_k))$ \\
			\quad \quad  \quad  -~Perform the recursion as:\\ 
			\begin{tabular}{p{8cm}}
				
				\tableequation 
					{\quad \quad\quad \quad  \quad \text{f}_{k+1}=(\textbf{I}_{N\times N}-\lambda\textbf S)\text{g}_k+\lambda \text{f}_s \label{eq:sol}} \\

			\end{tabular}\\
			\quad \quad  \quad   -~ ${\tilde{\text{f}}}=\text{f}_{k+1} $ \\
			\quad \quad  \quad   -~$k=k+1$ \\
			\quad\tabitem End While\\

			\hline
			
		\end{tabular}
		
	\end{table}
	
\section{CONVERGENCE ANALYSIS}
\label{sec:converge}	
In this section, we discuss convergence of the proposed IMATGI algorithm. To proceed, we need to prove the following lemma.

\textbf{Lemma 1}: Let\textquotesingle s denote the GFT of the sub-sampled graph signal by $\hat{\text{f}}_s=\textbf U^t\textbf S \text{f}$. Also, assume that the diagonal elements of the sampling matrix $\textbf{S}$ are independent identically distributed (iid) random variables coming from Bernoulli(p) distribution ($s_i\sim Bernoulli(p),\forall i$). We have:
\begin{align}
\label{eq:sol_2}
\quad\quad	\quad \quad E\{\hat {\text{f}}_s\} &= p\hat{\text{f}}\nonumber\\
E\{trace((\hat {\text{f}}_s-p\hat {\text{f}})(\hat {\text{f}}_s-p\hat {\text{f}})^t)\}&=(p-p^2)\epsilon
\end{align}
in which $\epsilon$ is the energy of the graph signal defined as $ \epsilon={\text{f}}^t {\text{f}}$.

\begin{proof} [\textbf{Proof:}\nopunct]
For the first equation we have:
   \begin{eqnarray}
   \label{eq:sol_3}
   E\{\hat {\text{f}}_s\}= E\{\mathbf{U^t}\mathbf{S}{\text{f}}\}=\mathbf U^tE\{\textbf{S}\}{\text{f}}=\mathbf{U^t}(p\textbf{I}){\text{f}}=p\mathbf{U^t}{\text{f}}=p\hat {\text{f}}  
   \end{eqnarray}
   
For the second equation we write (\ref{eq:sol_3}):
  \begin{eqnarray}
  \label{eq:sol_4}
 \hat {{\text{f}}_s}-p\hat {\text{f}}= \mathbf{U^t}\mathbf{S}{\text{f}}-p\mathbf{U^t}{\text{f}}=\mathbf{U^t}(\textbf{S}-p\textbf{I}){\text{f}}  
  \end{eqnarray}

  Now substituting (\ref{eq:sol_4}) we get:
  \begin{align}
  \label{eq:sol_5}
  &E\{trace((\hat {\text{f}}_s-p\hat {\text{f}})(\hat {\text{f}}_s-p\hat {\text{f}})^t)\}\nonumber\\&=E\{trace(\mathbf{U^t}(\mathbf {S}-p\textbf{I}){\text{f}}{\text{f}}^t(\mathbf {S}-p\textbf{I})^t\mathbf{U})\}\nonumber\\&=E\{trace((\mathbf {S}-p\textbf{I}){\text{f}}{\text{f}}^t(\mathbf {S}-p\textbf{I})^t)\}\nonumber\\&=trace(E\{(\mathbf {S}-p\textbf{I})^t(\mathbf {S}-p\textbf{I}){\text{f}}{\text{f}}^t\})
  \end{align}
  
   Now note that $\textbf{S}-p\textbf{I}$ is a diagonal matrix with $E\{(\textbf{S}-p\textbf{I})^t (\textbf{S}-p\textbf{I})\}=p(1-p)\textbf{I}$, hence we have:    	
    	\begin{align}
    	 \label{eq:sol_6}
    	 &E\{trace((\hat {\text{f}}_s-p\hat {\text{f}})(\hat {\text{f}}_s-p\hat {\text{f}})^t)\}\nonumber\\&=p(1-p) trace({\text{f}}{\text{f}}^t)=p(1-p)\epsilon 
    	\end{align}
    \end{proof}
\textbf{Theorem 1}: Under the assumptions of Lemma 1 (i.e. $s_i\sim Bernoulli(p),\forall i$), and considering the IMATGI reconstruction formula given in Table~\ref{tab:t1}, $\lim_{k\to\infty} \hat {\text{f}}_k$is an unbiased estimator of $\hat {\text{f}}$ for $0<\lambda<2/p$.
\begin{proof} [\textbf{Proof:}\nopunct]
To prove this theorem, we need to show that $E\{\lim_{k\to\infty}\hat {\text{f}}_k\}=\hat {\text{f}} $ or equivalently $\lim_{k\to\infty}E\{\hat {\text{f}}_k\}=\hat {\text{f}}$. ̂To this end, we define the sequence of error vectors $e_k=\hat {\text{f}}-E\{\hat {\text{f}}_k\}$ and show that each element of $e_k$ forms a geometric progression with common ratio $r=1-\lambda p$. Hence, if $0<\lambda<2/p$ the IMATGI reconstruction technique converges linearly (of order 1) to the original graph signal in the mean.

Starting the algorithm from a zero initial condition, we have $\hat {\text{f}}_0=0$ and hence $e_0=\hat {\text{f}}$. Also from the basic IMATGI recursion (\ref{eq:sol}) we have $\hat {\text{f}}_1=\lambda \hat {\text{f}}_s  $ and hence from Lemma 1 we get:
\begin{eqnarray}
\label{eq:sol_7}
 \text{e}_1=\hat {\text{f}}-E\{\hat {\text{f}}_1\}=(1-\lambda p)\hat {\text{f}}
\end{eqnarray}

Now note that the basic IMATGI recursion can be rewritten in transform domain as:
\begin{eqnarray}
\label{eq:sol_8}
\hat {\text{f}}_{k+1}=\textbf U^t(\textbf I_{N \times N}-\lambda \textbf S){\text{g}}_k+\lambda \hat {\text{f}}_s
\end{eqnarray}

Taking expected value from both sides of (\ref{eq:sol_8})  and utilizing Lemma 1 yields:
\begin{align}
\label{eq:sol_9}
E\{\hat {\text{f}}_{k+1}\}&=E\{\textbf U^t(\textbf I_{N \times N}-\lambda \textbf S){\text{g}}_k\}+\lambda p \hat {\text{f}} \nonumber\\&=\textbf U^t (1-\lambda p )\textbf I E\{\textbf U (\mathcal{T}(\textbf U^t {\text{f}}_k))\}+\lambda p \hat {\text{f}}\nonumber\\&=(1-\lambda p)E\{\mathcal{T}(\textbf U^t {\text{f}}_k)\}+\lambda p \hat {\text{f}}
\end{align}

Utilizing (\ref{eq:sol_9}) we get:

\begin{align}
\label{eq:sol_10}
\text{e}_{k+1}(i)&=\hat {\text{f}}-E\{\hat {\text{f}}_{k+1}\} \nonumber\\&=(1-\lambda p)\hat {\text{f}}-(1-\lambda p)E\{\mathcal{T}(\textbf U^t {\text{f}}_k)\}\nonumber\\&=(1-\lambda p)(\hat {\text{f}}-E\{\mathcal{T}(\hat {\text{f}}_k)\})
\end{align}

Now let\textquotesingle s take an elementwise look at the final equation (\ref{eq:sol_10}). Denote the $i$th element of the original signal, the estimated signal and the error vector by $\hat {\text{f}}(i),\hat {\text{f}}_k(i)$ and $e_{k+1} (i)$, respectively. We get (\ref{eq:sol_11}):
\begin{eqnarray}
\label{eq:sol_11}
\text{e}_{k+1}(i)=(1-\lambda p)(\hat {\text{f}}(i)-E\{\mathcal{T}(\hat {\text{f}}_k(i))\})
\end{eqnarray}

Now if $|\hat {\text{f}}_k(i)|\geqslant t(k)$, this element successfully passes the threshold. In this case we can omit the thresholding operator from the right side of (\ref{eq:sol_11}) and we get (\ref{eq:sol_12}):
\begin{align}
\label{eq:sol_12}
\text{e}_{k+1}(i)&=(1-\lambda p)(\hat {\text{f}}(i)-E\{\hat {\text{f}}_k(i)\})\nonumber\\&=(1-\lambda p)\text{e}_{k}(i)
\end{align}

On the other hand, if $|\hat {\text{f}}_k(i)|<t(k)$ then $\hat {\text{f}}_k(i)$ does not pass through the threshold and we have $E\{\mathcal{T}(\hat {\text{f}}_k)\}=0$ and it is obvious from (\ref{eq:sol_10}) that:
\begin{eqnarray}
\label{eq:sol_13}
\text{e}_{k+1}(i)=(1-\lambda p)\hat {\text{f}}(i)
\end{eqnarray}

Hence, once a vector element passes through the threshold in a specific iteration, its corresponding error sequence converges linearly to zero provided that $0<\lambda<2/p$. As the threshold is strictly decreasing and approaches zero as $k\rightarrow \infty$, all vector elements will eventually pass through the threshold and the proof is complete.
\end{proof}
In order to guarantee perfect reconstruction/convergence of the IMATGI algorithm, we also need to show that the variance of this unbiased estimator approaches zero as $k\rightarrow \infty$. Theorem 2 explains this variance fluctuation issue as $k$ approaches infinity. Before proceeding to the formal statement of Theorem 2, let’s define the support for the sparse graph signal $ {\text{f}}$ as the set of all non-zero elements in its GFT representation as $Supp=\{j|\hat  {\text{f}}(j)\neq 0\}$.

\textbf{Theorem 2}: Under the assumptions of Lemma 1 (i.e. $s_i\sim Bernoulli(p),\forall i$), if the GFT component $\hat  {\text{f}}_k(i) $ passes through the threshold in the $k$th iteration of the IMATGI algorithm, this decreases the estimation variance defined as $\sigma_k^2=E\{trace((\hat  {\text{f}}_k -E\{\hat  {\text{f}}_k\})(\hat  {\text{f}}_k-E\{\hat  {\text{f}}_k\})^t)\}$ if $i\in Supp$ and increases the variance for $i\not\in Supp$
\begin{proof} [\textbf{Proof:}\nopunct]
Let\textquotesingle s partition the set of all GFT components passed through the threshold at the $k$th iteration as $Supp_k= {\text{Q}}_k\bigcup  {\text{L}}_k$ in which $ {\text{Q}}_k$ represents the set of GFT components present in the original signal support (Supp) and $ {\text{L}}_k$ denotes the rest. Correspondingly, decompose $ {\text{g}}=\textbf U(\mathcal{T}(\textbf U^t  {\text{f}}_k ))$ as:
\begin{eqnarray}
\label{eq:sol_14} 	
 {\text{g}}_k= {\text{q}}_k+ {\text{l}}_k
\end{eqnarray}

In which $ {\text{q}}_k$ is the portion due to the support components and $ {\text{l}}_k$ is due to the non-support portion passed mistakenly through the threshold. Similarly, let\textquotesingle s decompose $ {\text{f}}$ as the sum of its reconstructed portion $ {\text{q}}_k$ and a residual $ {\text{r}}_k$ as 
\begin{eqnarray}
\label{eq:sol_15} 	
 {\text{f}}= {\text{q}}_k+ {\text{r}}_k
\end{eqnarray}

Now, substituting (\ref{eq:sol_14}) and (\ref{eq:sol_15}) in  (\ref{eq:sol}) gives:
\begin{align}
\label{eq:sol_16}
& {\text{f}}_{k+1}=(\textbf I_{N\times N}-\lambda \textbf S) {\text{g}}_k+\lambda  {\text{f}}_s\nonumber\\&=(\textbf I_{N\times N}-\lambda \textbf S)( {\text{q}}_k+ {\text{l}}_k)+\lambda \textbf{S}( {\text{q}}_k+ {\text{r}}_k)\nonumber\\&=\lambda \textbf{S}  {\text{r}}_k-\lambda \textbf{S} {\text{l}}_k+ {\text{q}}_k+ {\text{l}}_k
\end{align}

The last two terms in (\ref{eq:sol_16}) $( {\text{q}}_k \ and \   {\text{l}}_k)$ are not sub- sampled and hence do not contribute to the estimation variance $\sigma_{k+1}^2$. Utilizing Lemma 1, we can compute $\sigma_{k+1}^2$ due to the sub-sampled terms by (\ref{eq:lost}) 
\begin{align}
\label{eq:lost}
\sigma_{k+1}^2&=E\{trace((\hat  {\text{f}}_{k+1}-E\{\hat  {\text{f}}_{k+1}\})(\hat  {\text{f}}_{k+1}-E\{\hat  {\text{f}}_{k+1}\})^t)\}\nonumber\\&=\lambda^2(p-p^2)\epsilon_{ {\text{l}}_k}+\lambda^2(p-p^2)\epsilon_{ {\text{r}}_k}
\end{align}

In which $\epsilon_{ {\text{r}}_k}= {\text{r}}_k^t {\text{r}}_k$ and $\epsilon_{ {\text{l}}_k}= {\text{l}}_k^t {\text{l}}_k$ denote the energies of the residual and the portion due to the non-support components mistakenly passed through the threshold. As each mistakenly passed component $i\not\in Supp$ increases $\epsilon_{ {\text{l}}_k}$, it will consequently increase the spectrum variance. Similarly, for a correctly passed signal component $i\in Supp$, $\epsilon_{ {\text{r}}_k}$ and consequently the spectrum variance $\sigma_{k+1}^2$ is decreased. The above discussion completes the proof. 
\end{proof}
\textbf{Remark 1}: As stated previously, due to the non-zero spectrum variance, $\hat  {\text{f}}_k(i)$ is generally non-zero for $i\not\in Supp$. Hence, the threshold parameters must be adjusted such that the threshold value always keeps above the standard deviation at the $k$th iteration (e.g. $t(k)\geq \gamma\sigma_k$, $\gamma>1$) to prevent the algorithm from picking up incorrect GFT components. In this case, $\epsilon_{ {\text{l}}_k}=0$ and the estimation variance is decreasing in each iteration $\sigma^2_{k+1}\leq\sigma_k^2$.

\textbf{Corollary 1}: Considering Theorem 1, we conclude that the IMATGI estimation bias approaches zero as k approaches infinity. On the other hand, the variance of the IMATGI estimation is decreasing provided that the condition in Remark 1 $(\epsilon_{ {\text{l}}_k}=0,\forall k)$ always holds. Now considering the fact that the Mean Square Error (MSE) of the estimator is given by (\ref{eq:sol_17})
\begin{align}
\label{eq:sol_17}
&MSE_k=E\{trace((\hat  {\text{f}}_k-\hat  {\text{f}})(\hat  {\text{f}}_k-E\{\hat  {\text{f}}\})^t)\}\nonumber\\&=trace((E\{\hat  {\text{f}}_k\}-\hat  {\text{f}})(E\{\hat  {\text{f}}_k\}-\hat  {\text{f}})^t)+\sigma_k^2
\end{align}

As both terms in (\ref{eq:sol_17}) are decreasing, we conclude that $MSE_k$ is also decreasing. In other words, if the condition in Remark 1 holds, the IMATGI algorithm strictly decreases the Mean Square Error (MSE) of the estimated signal in each iteration. As MSE is a convex cost function, these sequential decreases converge to the unique global minimum.

Concluding this section, we have shown that perfect reconstruction by the proposed IMATGI algorithm is possible if the threshold parameters ($\alpha, \beta$) are selected such that the condition in remark 1 is satisfied, $0<\lambda<2/p$ and $k\rightarrow \infty$.
\section{Simulation results}
\label{sec:sim}
In this section we demonstrate efficient performance of the proposed IMATGI algorithm by simulations on both randomly generated sparse signals and three benchmark data sets used in recommendation systems.\\

\textbf{A. Generic Sparse Signals} \\

In order to fairly evaluate the performance of the proposed algorithm, we calculate and report the reconstruction SNR as (\ref{eq:sol_18}):
\begin{eqnarray} 
\label{eq:sol_18}
SNR=\frac{\vert\vert  {\text{f}}\vert\vert^2_2}{\vert\vert  {\text{f}}-\hat  {\text{f}}\vert\vert^2_2}
\end{eqnarray} 
where $ {\text{f}}$ and $\hat  {\text{f}}$ denote the original and reconstructed graph signals, respectively.

In this simulation scenario, we generate a graph with $N=1000$ randomly located nodes and edges similar to \cite{narang2013localized}. The signal entries $ {\text{f}}(i)$ associated with each node are taken from the uniform random variable $U(0,1)$. 
Now, define the number of sparse GFT components as $k$. In order to enforce sparsity of the generated signal in the GFT domain, we project the random signal onto the GFT domain $(\hat  {\text{f}}=\textbf{U}^t  {\text{f}})$, keep k entries with largest absolute values and set all the other GFT components to zero.

\begin{figure}[H]
	\centering
	\includegraphics[scale=.7]{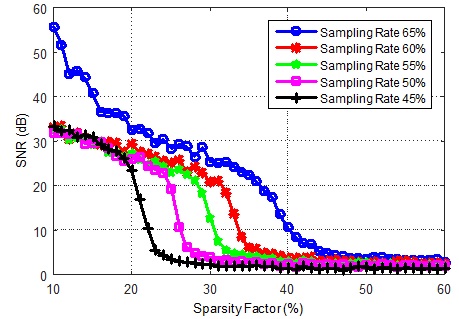}
		\caption{ The Reconstruction Performance for IMATGI}

	\label{fig:sysmodel31}
\end{figure}

To study the reconstruction performance of the proposed algorithm, we randomly sub-sample these generic $k$-sparse signals utilizing the sampling matrix $\textbf{S}$ at rates ranging from p=0.45 to p=0.65. We sweep the sparsity factor $\frac{k}{N}$ from 10\% to 60\%. For each sparsity factor and sampling rate, we repeatly generate 100 random $k$-sparse signals, sample randomly at rate $p$, reconstruct using 20 iterations of the proposed IMATGI algorithm and report the average achieved SNR in Fig.\ref{fig:sysmodel31}.

As observed in Fig.\ref{fig:sysmodel31}, all curves experience a sudden knee-like fall in reconstruction SNR as the sparsity factor increases. This fall is considered as the boundary between successful and unsuccessful reconstruction. As expected, the simulation results reveal that as the sampling rate increases, the algorithm can successfully reconstruct less sparse signals.

Note that as the number of iterations are bounded to 20 and the threshold parameters are not ideally trained in order to guarantee that Remark 1 holds, perfect reconstruction (infinite SNR value) is not observed in these simulations even on the left hand of the knee-like fall.\\

\textbf{B. Recommendation Systems } \\

In this scenario we compare the performance of the proposed IMATGI algorithm with the previously proposed graph interpolation methods in the widely desirable application of recommendation systems. To this end, we apply IMATGI on three benchmark datasets widely used for performance evaluations in recommendation systems \cite{WinNT,WinNT2,WinNT3}. To have a fair comparison between the performances of different methods, we report the normalized reconstruction RMSE values achieved (as defined by \cite{narang2013localized}) in Table~\ref{tab:t2}.

Following an approach similar to \cite{narang2013localized}, each dataset is reduced to a $100K$ randomly selected user-item sub-dataset and split into 5 fold cross-validation sets. In each iteration we use four subsets for training (i.e. is computing the graph and signal values) and the last subset for testing the performance of the algorithm \cite{narang2013localized}. 

Table~\ref{tab:t2} reports the RMSE values achieved by the proposed IMATGI algorithm along with the previously reported results for the other methods. As observed in this table, IMATGI improves the reconstruction performance in comparison with the literature. This is due to the fact that IMATGI utilizes the more general assumption of sparsity rather than bandlimitedness of the underlying graph signals. In fact, in this scenario, we observe that the real signals that arise in the application of recommendation systems are rather sparse than bandlimited (i.e. they have a few non-zero GFT components that may be located far apart from each other rather than condensed in a specific spectral range.) 

\section{Conclusion}

\label{sec:con}
In this paper we proposed the Iterative Method with Adaptive Thresholding for Graph Interpolation (IMATGI) algorithm for sparsity promoting interpolation of signals defined on graphs. We provided a formal convergence analysis for the proposed IMATGI algorithm and finally demonstrated its efficient reconstruction performance on both generic sparse data and the benchmark datasets for recommendation systems. 

\begin{table}[!h]
	\tiny
	\caption {RMSE Performance Comparison between Different Graph Interpolation Techniques for Recommendation Systems} \label{tab:t2}
	\centering
	\begin{tabular}{ |p{.85cm}|p{.65cm}|p{.65cm}|p{.65cm}|p{.65cm}|p{.65cm}|p{.65cm}|p{.65cm}| }
		\hline
	Dataset& KNN & PMF & RBM & IRBM & LSR & ILSR & IMATGI\\ \hline
	Movielens\cite{WinNT}  & 0.2482 & 0.2513 & 0.2414 & 0.2450 & 0.2514 & 0.2466 & \textbf{0.2406} \\ \hline
	Jester\cite{WinNT2} & 0.2348 & 0.2299 & 0.2304 & 0.2341 & 0.2344 &0.2315&\textbf{0.2130 } \\ \hline
	BX-books\cite{WinNT3} & 0.2677 &0.2093 & 0.1966 & 0.2138& 0.2651&0.2828&\textbf{0.1790} \\\hline
	\end{tabular}
\end{table}

\section{Acknowledgment}
Mahdi Boloursaz Mashhadi is supported by grants from Sharif University of Technology (SUT) and the Iranian National Science Foundation (INSF).
	
\bibliographystyle{ieeetr} 
\bibliography{ref}

\end{document}